\documentclass[
    ,final            % use final for the camera ready runs
  ]
  {aipproc}
\usepackage[usenames]{color}
\definecolor{red}{rgb}{1,0.0,0.}
\layoutstyle{6x9}
\newcommand{\coleps}{\epsilon_{I\alpha\beta}}
\newcommand{\flaeps}{\epsilon_{Iij}}
\newcommand{\setq}[2]{\{{\bf q}_{#1}^{#2}\}}

\newcommand{\intspace}[1]{\int d^4 {#1}\,}
\newcommand{\ha}{\frac{1}{2}}
\newcommand{\rr}{{\bf r}}
\newcommand{\q}[2]{ {\bf q}_{#1}^{#2}}
\newcommand{\vu}{{\bf u}}
\newcommand{\qia}{ {\bf q}_{I}^{a}}

\begin{document}
\title{The strength of crystalline color superconductors}
\author{Massimo Mannarelli}{
  address={Instituto de Ciencias del Espacio (IEEC/CSIC), E-08193
Bellaterra (Barcelona), Spain}}
\author{Krishna Rajagopal}{
  address={Center for Theoretical
Physics, MIT, Cambridge, MA 02139, USA}}

\author{Rishi Sharma}{
  address={Theoretical Division, Los Alamos National Laboratories, Los Alamos,
  NM 87545, USA}
}

\begin{abstract} 
We present a study of 
the shear modulus of the crystalline color superconducting
phase of quark matter, showing that this phase of dense, but not asymptotically
dense, quark matter responds to shear stress as a very rigid solid.  This phase
is characterized by a gap parameter $\Delta$ that is periodically modulated in
space and therefore spontaneously breaks translational invariance. We  derive
the effective action for the phonon fields that describe space- and
time-dependent fluctuations of the crystal structure formed by $\Delta$, and
obtain the shear modulus from the coefficients of the spatial derivative
terms.  Within a  Ginzburg-Landau approximation, we find shear moduli which
are 20 to 1000 times larger than those of neutron star crusts.  This 
phase of matter is thus more rigid 
than any known material in the universe, but at the same time
the crystalline color superconducting phase is also
superfluid. These properties raise the
possibility that the
presence of this phase
within neutron stars may have distinct
implications for their phenomenology. For example, (some) pulsar glitches may
originate in crystalline superconducting neutron star cores. 
\end{abstract}

\maketitle

%%%%%%%%%%%%%%%%%%%%%%%%%%%%%%%%%%%%%%%%%%%%
%% MAINMATTER
%%%%%%%%%%%%%%%%%%%%%%%%%%%%%%%%%%%%%%%%%%%%

\section{Introduction}
\vspace{-14cm} \hspace{12cm} MIT-CTP 3876 
\hspace*{11.8cm} LA-UR-07-6534 \vspace{12.5cm} \hspace{-12cm}

\noindent At the large baryon number 
densities found at the cores of
neutron stars, quarks may not
be confined into well-defined hadrons.
If quarks are deconfined, 
since the interaction between
quarks in the color antisymmetric channel is attractive, quarks near 
their Fermi
surfaces will tend to form Cooper
pairs.   Because the temperature of 
neutron stars is much smaller than the typical
pairing energy, if quark matter exists within neutron stars, it
must be in some color superconducting phase~\cite{reviews}. 

At asymptotically large densities where the masses of the $u$, $d$ and $s$
quarks can all be neglected, quark matter exists in the CFL
phase~\cite{Alford:1998mk} in which quarks of all three colors and all three
flavors form Cooper pairs with 
zero total momentum. The diquark condensate in the CFL phase 
is antisymmetric in color indices, 
driven by the attractive color interaction, and
antisymmetric in the spin indices. Thus it must be 
antisymmetric in flavor indices,
implying that quarks of a given flavor can only pair with quarks of 
the other two flavors. In the CFL phase,
all fermionic excitations are gapped, with a gap parameter $\Delta_0\sim 10-100$~MeV.

However, at densities relevant for neutron star phenomenology, meaning
quark chemical potentials at most $\mu\sim 500$ MeV, the strange quark mass $M_s$
cannot be neglected. In neutral unpaired quark matter in weak equilibrium, $M_s$ induces 
splitting between the Fermi surfaces for quarks of different flavor, which
can be taken into account to lowest order in $M_s^2/\mu^2$
by treating the quarks as if they were massless but with chemical
potential splittings $\delta\mu_2\equiv (\mu_u-\mu_s)/2$ and 
$\delta\mu_3\equiv (\mu_d -\mu_u)/2$ given 
by $\delta\mu_2=\delta\mu_3\equiv \delta\mu = M_s^2/(8\mu)$.
Note that the splitting between unpaired Fermi surfaces increases
with decreasing density. 
In the CFL phase, the Fermi momenta are {\it not} given by these optimal
values for unpaired quark matter; instead, the system pays a free energy price 
$\propto \delta\mu^2 \mu^2$ to equalize all Fermi momenta and gains a pairing energy benefit
$\propto \Delta_0^2\mu^2$.   As a function of decreasing density, there
comes a point (at which $\delta\mu \approx \Delta_0/4$~\cite{Alford:2003fq}) when
the system can lower its energy by breaking pairs. Restricting the analysis to spatially homogeneous condensates, 
the phase that results when CFL Cooper pairs start to break is the gapless CFL  (gCFL)
phase~\cite{Alford:2003fq}. However, this phase turns out to be 
``magnetically unstable'' \cite{Casalbuoni:2004tb}, meaning that it is unstable to 
the formation of counter-propagating currents. If $\Delta_0$ is small enough
that  CFL pairing cannot survive all the way down to
the $\mu$
at which quark matter is supplanted by nuclear matter, then
the true ground state of intermediate density quark matter 
must have 
a lower free energy than that of the unstable gCFL phase.

Crystalline color superconducting quark matter is a possible resolution
of the magnetic instability of the gCFL phase.
Crystalline color 
superconductivity~\cite{Alford:2000ze,Bowers:2002xr,Mannarelli:2006fy,Rajagopal:2006ig} 
is the QCD analogue of a form of non-BCS pairing
first considered by Larkin, Ovchinnikov, Fulde and Ferrell~\cite{LOFF} . 
This phase may be the ground state of matter
in the intermediate density regime in which quark matter is favored
over nuclear matter but the Fermi surface separations $\propto M_s^2/\mu$
are large enough to disrupt CFL pairing.  In a crystalline color superconducting
phase,
quarks whose Fermi surfaces
are separated (as favored in the absence of pairing)
nevertheless pair. Unlike in conventional BCS
phases, the quarks in a pair do not have equal and opposite
momenta (meaning that the Cooper pairs have net momentum)
allowing both members in a pair to have momenta near their
respective, separated, Fermi surfaces.
Such phases do not suffer
from the magnetic instability~\cite{Ciminale:2006sm}.
A particular crystalline phase is specified by sets
of momentum vectors $\setq{I}{}$, meaning that
Cooper pairs with total momentum $2{\bf q}_I^a$ form
for each ${\bf q}_I^a \in \setq{I}{}$.
All the ${\bf q}_I^a$'s have the same magnitude
$q_I\equiv |{\bf q}_I^a |=\eta\delta\mu_I$, with
$\eta=1.1997$~\cite{Bowers:2002xr} and
$2\delta\mu_I$ the separation between the Fermi surfaces of quarks that pair.
The directions of the vectors in $\setq{I}{}$
must be determined to get structures with the smallest free energy.
In position space, the condensate is
\begin{equation}
\langle\psi_{i\alpha}C\gamma^5\psi_{j\beta}\rangle\propto
\sum_I\coleps\flaeps\Delta_I\sum_{{\bf q}_I^a\in \setq{I}{}}\exp({2i{\bf q}_I^a\cdot{\bf r}})\ .
\label{condensate}
\end{equation}
This is antisymmetric in color ($\alpha,\beta$), spin, 
and flavor ($i,j$) (where ($1$, $2$, $3$) correspond to ($u$, $d$, $s$)
respectively) indices and is thus a generalization
of the CFL condensate to crystalline color superconductivity. 
For simplicity, $\Delta_1$ is set to $0$, neglecting
$\langle ds \rangle$ pairing because the $d$ and $s$ Fermi
surfaces  are twice as far apart from each other as each is from
the intervening $u$ Fermi surface. Hence, $I$ can be taken to run over $2$ and $3$ only.
$\setq{2}{}$ and $\setq{3}{}$ define the crystal structures of the $\langle us \rangle$ and
$\langle ud \rangle$ condensates respectively.

We will analyze crystalline color superconductivity in an NJL model, which gives in the mean field
approximation an interaction term
\begin{equation}
{\cal L}_{\rm interaction}= \frac{1}{2} \bar{\psi}\Delta({\bf r})\bar{\psi}^T +
h.c.,
\label{meanfieldapprox}
\end{equation}
where the proportionality constant in Eq.~(\ref{condensate}) is conventionally
chosen so that,
\begin{equation}
\Delta(\rr) = (C\gamma^5)
\sum_I\coleps\flaeps\Delta_I\sum_{{\bf q}_I^a\in\setq{I}{}}\exp({2i{\bf q}_I^a\cdot{\bf r}})
\label{precisecondensate}\;.
\end{equation}

The authors of~\cite{Rajagopal:2006ig} calculated the free energy $\Omega$ of
several crystalline structures within the weak coupling ($\delta\mu, \Delta_0\ll \mu$)
and Ginzburg-Landau ($\Delta\ll\delta\mu,\Delta_0$)
approximations, and found qualitative features that make a structure
favorable. Two particular structures that possess these features, called CubeX
and 2Cube45z and described below, have a lower $\Omega$ than 
any other crystal patterns that have been 
analyzed. One or the other or both of these two is favored over unpaired quark matter and the
gapless CFL phase over the large range of densities given by~\cite{Rajagopal:2006ig}
\begin{equation}
2.9\Delta_0<\frac{M_s^2}{\mu}<10.4\Delta_0\;.
\end{equation}
For $\Delta_0=25$MeV and $M_s=250$MeV, the 
window of densities over which one of these crystalline phases has
the lowest free energy of any color superconducting phase extends
from well below the $\mu$ at which nuclear matter supersedes quark
matter to well above the highest $\mu\sim 500$~MeV expected in the
cores of neutron stars. The robustness of these phases is partly due to
their having reasonably large 
gap parameters $\Delta$, so large that the Ginzburg-Landau expansion parameter
$(\Delta/\delta\mu)^2$ can be around a tenth to a fourth meaning that this
approximation is at the edge of its validity.
Nevertheless, 
their impressive robustness over a large range of
$\mu$ relevant for cores of neutron stars make it worth 
considering the  phenomenological implications of their presence.  
We also note that in the case of a simpler crystal pattern, for which
results have been obtained without making the Ginzburg-Landau
approximation, this approximation is conservative in that it always
underestimates both $\Delta$ and the condensation energy~\cite{Mannarelli:2006fy}.

In the CubeX crystal structure, 
$\setq{2}{}$ and $\setq{3}{}$ each contain
four unit vectors, with
$\{{\bf \hat q_2}\} = \{({1}/{\sqrt{3}}) (\pm \sqrt{2},0,\pm1)\}$ 
and $\{{\bf \hat q_3}\} = \{({1}{\sqrt{3}}) (0,\pm\sqrt{2},\pm1)\}$. 
In the 2Cube45z crystal structure
$\setq{2}{}$ and $\setq{3}{}$ each contain
eight unit vectors,
with $\{{\bf \hat q_2}\} = \{(1/\sqrt{3})(\pm 1,\pm 1,\pm 1)\}$  and  
$\{{\bf \hat q_3}\} = \{(1/\sqrt{3})(\pm \sqrt{2},0,\pm 1)\}\cup\{(1/\sqrt{3})(0,\pm
\sqrt{2},\pm 1)\}$. For these structures, $\{{\bf \hat q_2}\}$ can be exchanged
with $\{{\bf \hat q_3}\}$ by rigid rotations, ensuring that there are electrically neutral
solutions of the gap equation with
$\Delta_2=\Delta_3=\Delta$~\cite{Rajagopal:2006ig}, a fact that we will
use in the next Section.

\section{Phonons}

The crystalline phases of color superconducting quark matter that we have
described in the previous Section are unique among all forms of dense matter
that may arise within neutron star cores in one respect: they are
rigid~\cite{Mannarelli:2007bs}.  They are not solids in the usual sense: the
quarks are not fixed in place at the vertices of some crystal structure.
Instead, these phases are in fact superfluid since the condensates all
spontaneously break the $U(1)_B$ symmetry corresponding to quark number.  
The diquark condensate, although spatially inhomogeneous,
can carry supercurrents~\cite{Alford:2000ze,Mannarelli:2007bs}. 
And yet, we shall see that crystalline color superconductors are rigid solids with
large shear moduli.  It is the
spatial modulation of the gap parameter that breaks translation invariance, 
and it is this
pattern of modulation that is rigid.
This novel form of rigidity may sound tenuous
upon first hearing, but we shall present the effective Lagrangian that describes
the phonons in the CubeX and 2Cube45z crystalline phases, whose lowest order
coefficients have  been calculated in the NJL model that we are
employing~\cite{Mannarelli:2007bs}.  We shall then extract the shear moduli from
the phonon effective action, quantifying the rigidity and indicating the
presence of transverse phonons.  

The shear moduli of a crystal may be extracted from the effective Lagrangian
that describes phonons in the crystal, namely space- and time-varying displacements
of the crystalline pattern~\cite{Casalbuoni:2002my}.  
 In the present context, we introduce displacement fields
for 
the $\langle ud \rangle$,
$\langle us \rangle$ and $\langle ds \rangle$  condensates 
by making the replacement
\begin{equation}
\Delta_I \sum_{\q{I}{a}\in\setq{I}{}}e^{2i\q{I}{a}\cdot\rr} \rightarrow
\Delta_I \sum_{\q{I}{a}\in\setq{I}{}}e^{2i\q{I}{a}\cdot(\rr - \vu_I(\rr))}
\label{displacementfields}
\end{equation}
in (\ref{precisecondensate}).  
One way to obtain the effective action describing the dynamics of the
displacement fields $\vu_I(\rr)$, 
including both its form and the values of its coefficients
within the NJL model that we are employing, is to take the mean field NJL
interaction given by Eq.~(\ref{meanfieldapprox}), but with
(\ref{displacementfields}), 
and integrate out the fermion fields. Since the gapless fermions do not
contribute to the shear modulus, one can integrate out the fermions completely
for this calculation. As an aside, 
we mention that this is not true for the calculation of thermal or
transport properties, where the gapless fermions do contribute and indeed
may dominate, as is the case for the heat capacity~\cite{Casalbuoni:2003sa} 
and neutrino 
emissivity~\cite{Anglani:2006br}.

After integrating out the fermions, we obtain~\cite{Mannarelli:2007bs}
\begin{eqnarray}
S[{\bf u}]&=&%&
\ha\intspace{x}\sum_I \kappa_I\nonumber
\times\Biggl[
  \left(
	\sum_{\qia\in\setq{I}{}}(\hat{q}_I^a)^m(\hat{q}_I^a)^n \right)
	(\partial_0 u_I^m)(\partial_0 u_I^n) \Biggr.\\
& & \Biggl.-\left(
\sum_{\qia\in\setq{I}{}}(\hat{q}_I^a)^m(\hat{q}_I^a)^v(\hat{q}_I^a)^n(\hat{q}_I^a)^w\right)
	(\partial_v u_I^m)(\partial_w u_I^n)
 \Biggr] \label{Seff4}\;
\end{eqnarray}
where $m$, $n$, $v$ and $w$ are spatial indices running over $x$, $y$ and $z$
and where we have defined
$ \kappa_I\equiv
\frac{2\mu^2|\Delta_I|^2\eta^2}{\pi^2(\eta^2-1)}$.  For 
 $\Delta_1=0,\;\;\Delta_2=\Delta_3=\Delta$, and $\eta\simeq1.1997$,
$\kappa_2=\kappa_3\equiv\kappa\simeq 0.664\,\mu^2|\Delta^2|.$
$S[{\bf u}]$  is the low energy effective action for phonons in any crystalline color
superconducting phase, valid to second order in derivatives, to
second order in the gap parameters $\Delta_I$ and to second order in
the phonon fields $\vu_I$.    Because we are interested in long wavelength,
small amplitude, phonon excitations, expanding to second order in derivatives
and in the phonon fields is satisfactory. The Ginzburg-Landau approximation is
an expansion in $(\Delta/\delta\mu)^2$, and so is not under quantitative control for
the most favorable phases, as we have discussed.
But, the main 
requirement from glitch phenomenology for
the shear modulus is that it should be large, and
given that we find {\it much} larger values than those obtained for conventional
neutron star crusts, there is no great motivation to 
go to higher
order to improve the precision. (A higher order calculation would include
couplings between the different $\vu_I$, meaning they could no longer
be treated independently.)

According to the theory of elastic media~\cite{Landau:Elastic}, the  shear moduli can be extracted from  the phonon
effective action. Introducing the strain tensor
\begin{equation}
s_I^{mv}\equiv\ha\Bigl(\frac{\partial  u_I^m}{\partial
x^v}+\frac{\partial  u_I^v}{\partial x^m}\Bigr)\label{strain}\,,
\end{equation}
we then wish to compare the action (\ref{Seff4}) to
\begin{equation}
{S}[{\bf u}]= \ha\intspace{x}\Biggl(
	 \sum_I\sum_m \rho_I^m (\partial_0  u_I^m)(\partial_0  u_I^m)
	 -\sum_{I}\sum_{{mn}\atop{vw}}\lambda_I^{mvnw}
	 s_I^{mv}s_I^{nw}\Biggr)
\label{full action},
\end{equation}
which is the general form of the action quadratic in displacement fields
and which defines the elastic modulus tensor $\lambda_I^{mvnw}$.
In this case, the stress tensor 
(in general the derivative of the potential energy with respect to $s_I^{mv}$) is given by
\begin{equation}
\sigma_I^{mv}
=
\lambda_I^{mvnw}s_I^{nw}\label{stress2}\; .
\end{equation}
The diagonal components of $\sigma$ are proportional to the
compression exerted on the system and are therefore related to  the
bulk  modulus of the crystalline color superconducting
quark matter. Since unpaired quark
matter  has a pressure $\sim \mu^4$, it gives a contribution to the
bulk modulus that  completely overwhelms the contribution from the
condensation into a crystalline phase, which is of order
$\mu^2\Delta^2$.  We shall therefore not calculate the
bulk modulus.  On the other hand, the response to shear
stress arises only because of the presence  of the crystalline condensate.
The shear modulus is defined as follows. Imagine exerting a
static external stress $\sigma_I$ having only an off-diagonal
component, meaning  $\sigma^{mv}_I\neq 0$ for a pair of space
directions $m\neq v$, and all the other components of $\sigma$ are
zero. The system will respond with a strain $s_I^{nw}$. 
The shear modulus in the $mv$ plane is then
\begin{equation}
\nu_I^{mv} \equiv \frac{\sigma_I^{mv}}{2s_I^{mv}} 
= \ha\lambda_I^{mvmv}
\label{defineshearmodulus}\;,
\end{equation}
where the indices $m$ and $v$ are not summed.  
For a general quadratic potential
with $\sigma_I^{mv}$ given by (\ref{stress2}), $\nu_I^{mv}$ simplifies
partially but the full simplification given by the last equality in  
(\ref{defineshearmodulus}) only arises for special cases in which 
the only nonzero entries in $\lambda^{mvnw}$ with $m\neq v$ are
the $\lambda^{mvmv}$ entries, as is the case for all the crystal
structures that we consider.

For a given crystal structure, upon evaluating the sums in (\ref{Seff4}) and
then using the definition (\ref{strain}) to compare (\ref{Seff4}) to
(\ref{full action}), we can extract expressions for the $\lambda$ tensor
and thence for the shear moduli.  
This analysis, described in detail in \cite{Mannarelli:2007bs},
shows that in the CubeX phase
\begin{equation}
\nu_2=\frac{16}{9}\kappa\left( \begin{array}{ccc}
0 & 0 & 1\\
0 & 0 & 0\\
1 & 0 & 0
\end{array}
\right)\,,\hspace{.3cm}  \nu_3=\frac{16}{9}\kappa\left(
\begin{array}{ccc}
0 & 0 & 0\\
0 & 0 & 1\\
0 & 1 & 0
\end{array}
\right)\label{nu2 and nu3}\;,
\end{equation}
while in the 2Cube45z phase
\begin{equation}
\nu_{2}=\frac{16}{9}\kappa\left( \begin{array}{ccc}
0 & 1 & 1\\
1 & 0 & 1\\
1 & 1 & 0
\end{array}
\right)\,,\ \
\nu_{3}=\frac{16}{9}\kappa\left( \begin{array}{ccc}
0 & 0 & 1\\
0 & 0 & 1\\
1 & 1 & 0
\end{array}
\right)\label{nu 2Cube45z}\;.
\end{equation}
We shall see in the next Section that it is relevant to check
that both these crystals have enough 
nonzero entries in their shear moduli
$\nu_I$ that if there are rotational vortices are pinned within them, 
a force seeking to 
move such a vortex is opposed by the rigidity 
of the crystal structure described
by one or more of the nonzero entries in the $\nu_I$.  
This is demonstrated
in \cite{Mannarelli:2007bs}.

We see that all the nonzero shear moduli of both the CubeX and 2Cube45z
crystalline color superconducting phases turn out to take on the same value,
\begin{equation}
\nu_{\rm CQM} = \frac{16}{9}\kappa
=  1.18\, \mu^2 \Delta^2
=  2.47\, \frac{{\rm MeV}}{{\rm fm}^3}
\left(\frac{\Delta}{10~{\rm MeV}}\right)^2 \left(\frac{\mu}{400~\rm{MeV}}\right)^2,
\label{shearmodulus}
\end{equation}
where $\mu$ is expected to  lie between $350$ and $500$MeV and $\Delta$ may be
taken to lie  between $5$ and $25$MeV to obtain numerical estimates.

 From (\ref{shearmodulus}) we first of all see that the shear modulus
is in no way suppressed relative to the scale  $\mu^2\Delta^2$ that could have
been guessed on dimensional grounds.  And, second, we discover that
a quark matter core in a crystalline color superconducting phase is 20 to 1000
times more rigid than the crust of a conventional neutron 
star~\cite{Strohmayer:1991}.
Finally,  one can extract the phonon dispersion relations from the effective 
action~(\ref{Seff4}).
The transverse phonons, whose restoring force is provided by the shear 
modulus turn out to have direction-dependent velocities that are 
typically a substantial
fraction of the speed of light, %in the specific instances evaluated in
being given by $\sqrt{1/3}$ and $\sqrt{2/3}$
\cite{Mannarelli:2007bs}.
This is another way of seeing that this 
superfluid phase of matter is rigid indeed.

\section{Rigid quark matter and pulsar glitches}
%\label{sec:astro_glitches}

The existence of a rigid crystalline color superconducting core within
neutron stars may have a variety of observable consequences.   
For example,
if some agency (like magnetic
fields not aligned with the rotation axis) could maintain the
rigid core in a shape that has a nonzero quadrupole moment, gravity 
waves would be emitted.  
The 
LIGO non-detection of such gravity waves
from nearby neutron stars~\cite{Abbott:2007ce} 
already limits the possibility that they have rigid cores
that are deformed to the maximum extent allowed by the shear
modulus (\ref{shearmodulus}),
upon assuming a range of breaking strains,
and this constraint will tighten as
LIGO continues to run~\cite{Haskell:2007sh}. 
Perhaps the most exciting implication of a rigid core, however,
is the possibility that (some) pulsar
``glitches'' could originate deep within a neutron star, in its quark matter core. 

A spinning neutron star observed
as a  pulsar gradually spins down as it loses rotational energy
to electromagnetic radiation.
But, every once in a while the angular velocity
at the crust of the star is observed to increase suddenly in a dramatic event called a 
glitch.
The standard 
explanation~\cite{Anderson:1975} 
requires the presence of
a superfluid in some region of the star which also
features a rigid structure that 
can pin the vortices in the rotating superfluid and that does not easily deform
when the vortices pinned to it are under tension.
As a spinning pulsar slowly loses angular momentum over years,
since the angular momentum of any
superfluid component of
the star is proportional to the density of vortices,
the vortices ``want'' to move apart.
However, if  
some vortices are pinned to a rigid structure and so do not move, 
after a time 
this superfluid component of the star is spinning faster
than the rest of the star.  When the ``tension'' built up
in the array of pinned vortices
reaches a critical value, there
is a sudden ``avalanche'' in which vortices unpin, move outwards 
reducing the angular momentum of the superfluid, {\it and then re-pin}.  
As this superfluid suddenly loses angular momentum, the rest 
of the star, including in particular the surface whose angular
velocity is observed, speeds up --- a glitch.   
In the standard explanation of pulsar glitches, the necessary
conditions are met in the inner crust of a neutron star
which features a neutron superfluid coexisting with
a rigid array of positively
charged nuclei that may serve as vortex pinning sites.  
In very recent work, Link has questioned whether this scenario
is viable
because once neutron vortices
are moving through the inner crust, as must happen during a glitch,
they  are so resistant to bending
that they  may never re-pin~\cite{LinkPrivateCommunication}.  
Link concludes that we do not have an understanding
of any dynamics that could lead to the re-pinning of moving
vortices in the crust, and hence that we do not currently
understand the origin of glitches as a crustal phenomenon.

By virtue of being simultaneously superfluids
and rigid solids, the crystalline
phases of quark matter provide all the necessary conditions
to be the locus in which (some) pulsar glitches
originate.     
Their shear moduli
(\ref{shearmodulus}) 
make them more than rigid enough for glitches to originate within them.
The crystalline phases are at the same time superfluid, and
it is reasonable to expect that the superfluid vortices
that result when a neutron star with such a core rotates
 have lower free energy if they are centered
along the intersections of the nodal planes of the underlying
crystal structure, i.e. along lines along which the condensate
already vanishes  in the absence of a rotational vortex.
A crude estimate of the pinning force on vortices
within crystalline color superconducting quark matter indicates
that it is sufficient~\cite{Mannarelli:2007bs}.
So, the basic requirements for superfluid vortices pinning to
a rigid structure are all present.
The central questions that remain to be addressed are
the explicit construction of vortices in the crystalline phase
and the calculation of their pinning force, as well
as the calculation of the timescale over which sudden changes in the
angular momentum of the core are communicated
to the (observed) surface, presumably via
the common electron fluid or via magnetic stresses.

Much theoretical work remains before the hypothesis
that pulsar glitches originate within a crystalline color superconducting
neutron star core is developed fully enough to allow it to
confront data on the magnitudes, relaxation
timescales, and repeat rates that characterize glitches.  Nevertheless,
this hypothesis offers one immediate advantage over the conventional
scenario that relied on vortex pinning in the neutron star crust.  
It is impossible for a neutron
star anywhere within which rotational vortices 
are pinned to precess on $\sim$ year time scales~\cite{Link:2006nc},
and yet there is now evidence that
several pulsars are precessing~\cite{Stairs:2000}.
Since {\it all} neutron stars have crusts, the precession
of any pulsar is inconsistent with the pinning of vortices
within the crust, a requirement in the standard explanation of
glitches.  On the other hand, perhaps not all neutron stars have crystalline
quark matter cores --- 
for example, perhaps the lightest
neutron stars have nuclear matter cores.
Then, if vortices are never
pinned in the crust but are pinned within
a crystalline quark matter core,
those neutron stars that do have a crystalline quark matter core can glitch
but cannot precess while those that don't can precess but cannot glitch.

%%%%%%%%%%%%%%%%%%%%%%%%%%%%%%%%%%%%%%%%%%%%%%%%
%% BACKMATTER
%%%%%%%%%%%%%%%%%%%%%%%%%%%%%%%%%%%%%%%%%%%%%%%%
\begin{theacknowledgments}
The work of MM has been supported
by the ``Bruno Rossi" fellowship program.   This research
was supported in part by the Office of Nuclear Physics of the Office
of Science of the U.S.~DOE under grants
\#DE-AC02-05CH11231 and
\#DE-FG02-94ER40818 and LANS, LLC for the NNSA 
of the DOE under contract \#DE-AC52-06NA25396.
\end{theacknowledgments}
\vspace*{-.1cm}

\bibliographystyle{aipproc}   % if natbib is available

\end{document}